\documentclass[rn,11pt]{ucl_document} %http://mediatools.cs.ucl.ac.uk/nets/dos/b
\usepackage{graphicx}
\usepackage{hyperref} %/shareopt/tetex/share/texmf/tex/latex/hyperref/hyperref.sty

\newlength{\halfwidth}
\newlength{\temp}

\begin{document}

\title{Human-Competitive Awards 2018}

\date{22 October 2018}
\documentnumber{18/07}

\author{
{W. B.~Langdon}
}

\maketitle

\pagenumbering{arabic}  %turn on page numbering cf sig-alt-release2.cls
\pagestyle{plain}

\vfill
\centerline{To appear in \href{http://www.sigevolution.org/}{SIGEVOlution}}
\vfill

\newpage
\vfill
\section{Human-Competitive Awards 2018: The ``Humies''}
\vfill

 %make sure prize winners appear in references in the correct order
\nocite{Lones:2017:JMS}
\nocite{Kelly:2018:EC}
\nocite{Hart:2017:GECCO}
\nocite{Ceska:2017:ICCAD}

\noindent
The 
\href{http://gecco-2018.sigevo.org/}
{GECCO 2018} conference in Kyoto, Japan
hosted the $15^{\rm th}$ annual ``Humies'' Awards.

The first annual ``Humies'' competition was held at the 
2004 Genetic and Evolutionary Computation Conference (GECCO-2004)
in Seattle (USA).
With its generous prize money
(provided by John Koza)
it has become a staple of the Genetic and Evolutionary Computing calendar.
The Humies offer the opportunity to the EC community to showcase its
best work,
work, which by definition,
is better than human.

\vfill
\begin{figure}[!hb]
\centering
\includegraphics[width=\textwidth]{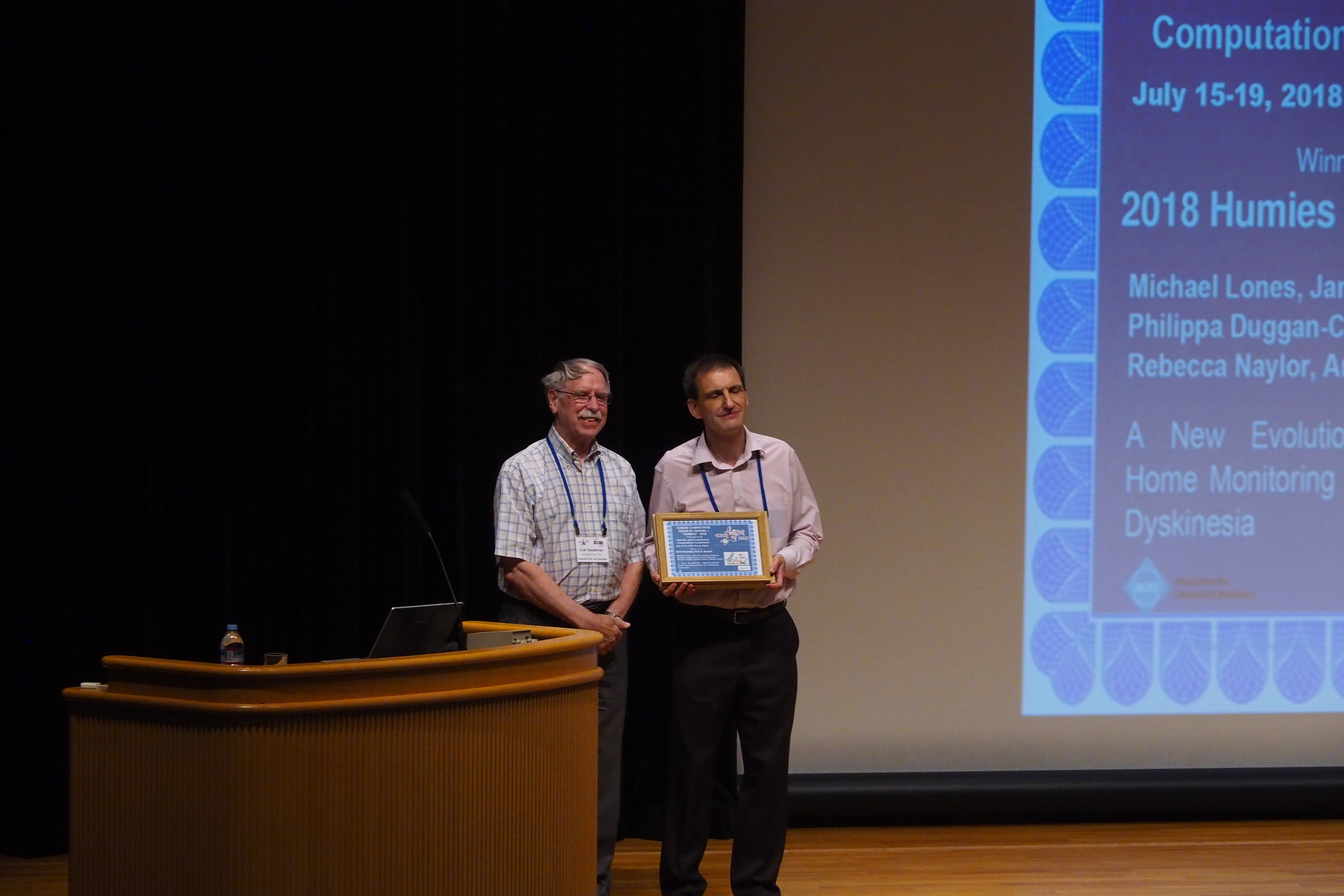}
\caption{Erik Goodman
presenting the first prize (and \$5\,000) to 
\protect\href{http://www-users.york.ac.uk/~sls5/}
{Steve Smith}
for 
``A New Evolutionary Algorithm-Based Home Monitoring Device for Parkinson's Dyskinesia''
Journal of Medical Systems (2017)
\protect\cite{Lones:2017:JMS}.
(Photo from Zdenek Vasicek.
More photos at 
\href{http://www.cs.ucl.ac.uk/staff/W.Langdon/gecco2018/\#humies}
     {http://\allowbreak{}www.cs.ucl.ac.uk/\allowbreak{}staff/\allowbreak{}W.Langdon/\allowbreak{}gecco2018}).
}
\label{fig:gold}
\end{figure}
\vfill

\newpage
\section{What it means to be Human-Competitive: Eight Criteria}

Although the judges will eventually have to deliberate and then decide,
to run a competition we need to be reasonably precise
by what we mean by ``Human-Competitive''.
So to enter the Humie competition it is necessary to show that you
satisfy one
of the following 
8~criteria, which where originally proposed by Koza
in the $10^{\rm th}$ anniversary issue of GP+EM%
~\cite{Koza:2010:GPEM}%
\footnote{We are now approaching the $20^{\rm th}$ anniversary
of ``Genetic Programming and Evolvable Machines''
and there will be a 
\href{http://gpemjournal.blogspot.com/2018/07/cfp-gpem-20th-anniversary-issue.html}
{special issue} to celebrate.}:

\begin{itemize}
\item 
(A) The result was patented as an invention in the past, is an improvement over a patented invention, or would qualify today as a patentable new invention.

\item 
(B) The result is equal to or better than a result that was accepted as a new scientific result at the time when it was published in a peer-reviewed scientific journal.

\item 
(C) The result is equal to or better than a result that was placed into a database or archive of results maintained by an internationally recognized panel of scientific experts.

\item 
(D) The result is publishable in its own right as a new scientific result — independent of the fact that the result was mechanically created.

\item 
(E) The result is equal to or better than the most recent human-created solution to a long-standing problem for which there has been a succession of increasingly better human-created solutions.

\item 
(F) The result is equal to or better than a result that was considered an achievement in its field at the time it was first discovered.

\item 
(G) The result solves a problem of indisputable difficulty in its field.

\item 
(H) The result holds its own or wins a regulated competition involving human contestants (in the form of either live human players or human-written computer programs).
\end{itemize}

\vfill
\section{Judges}

\setlength{\temp}{\textwidth}
\addtolength{\temp}{-8\tabcolsep}
\setlength{\temp}{0.2\temp}
\begin{tabular}{@{}ccccc@{}}
\includegraphics[width=\temp]{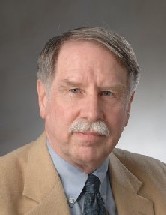}&
\includegraphics[width=\temp]{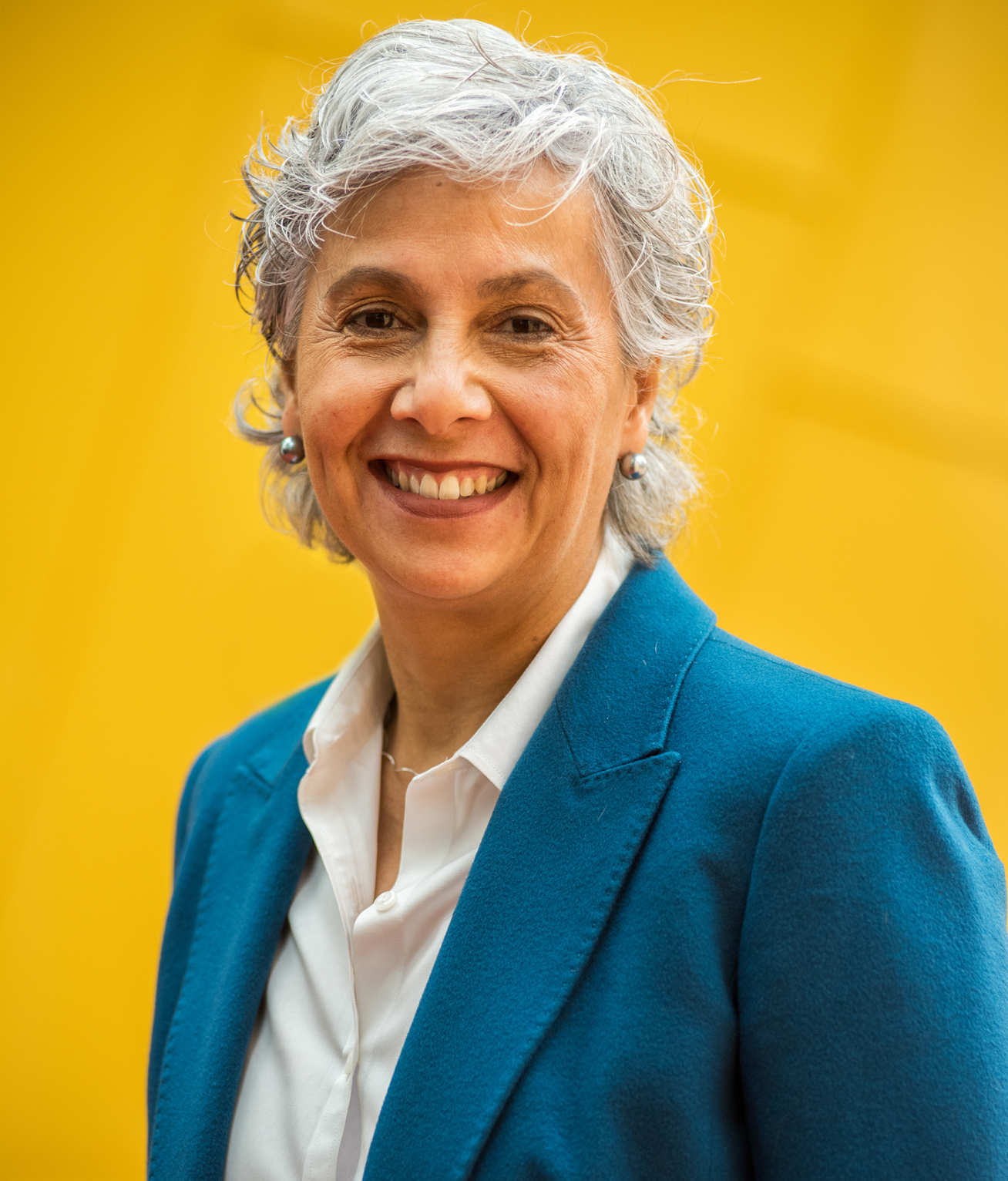}&
\includegraphics[width=\temp]{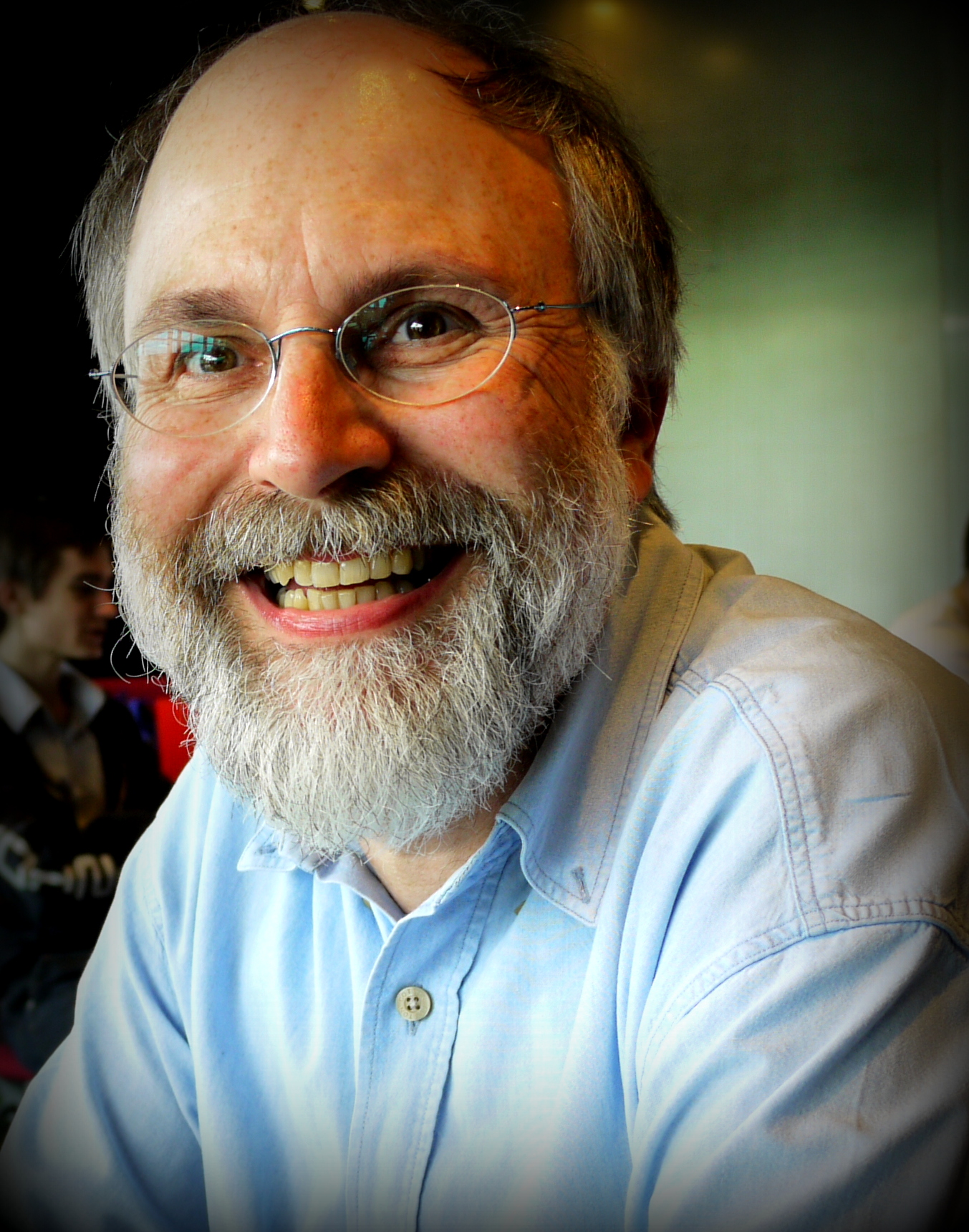}&
\includegraphics[width=\temp]{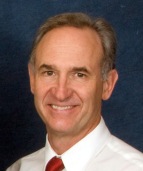}&
\includegraphics[width=\temp]{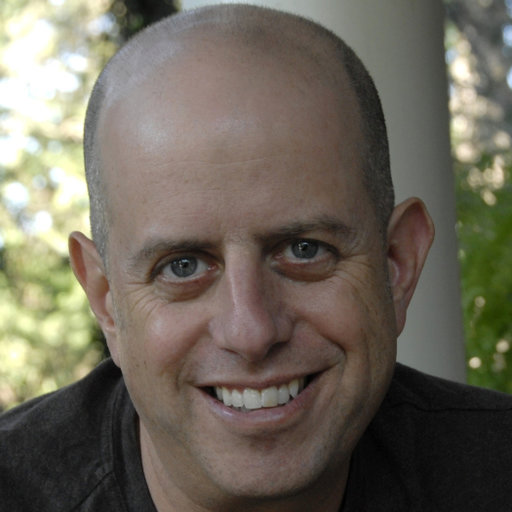}
\\
\href{http://www.egr.msu.edu/~goodman/}{Erik Goodman}&
\href{http://people.csail.mit.edu/unamay/}{Una-May O'Reilly}&
\href{http://www.cse.msu.edu/~banzhafw/}{Wolfgang Banzhaf}&
\href{http://www.cs.colostate.edu/~whitley/}{Darrell Whitley}&
\href{http://hampshire.edu/lspector/}{Lee Spector}
\end{tabular}

\vfill
\newpage
\section{Results}

This year there were 16 entries.
The judges selected nine to be finalists.
The finals where held in front of the judges at this year's GECCO
conference.
As a result they awarded 
\begin{itemize}
\item First prize: the Gold Award, \$5\,000,
to
\begin{itemize}
\item 
Michael Lones        (Heriot-Watt University, Edinburgh, UK),
\item 
Jane Alty            (Consultant in Neurology, Leeds Teaching Hospitals),
\item 
Jeremy Cosgrove      (Leeds Teaching Hospitals NHS Trust),
\item 
Philippa Duggan-Carter (Department of Neurology, Leeds General Infirmary),
\item 
Stuart Jamieson      (Department of Neurology, Leeds General Infirmary),
\item 
Rebecca Naylor       (York University),
\item 
Andrew Turner        (York University),
and
\item 
Stephen Smith        (York University)
\end{itemize}
for
``A New Evolutionary Algorithm-Based Home Monitoring Device for Parkinson's Dyskinesia''
Journal of Medical Systems (2017)
\cite{Lones:2017:JMS}

\item Second prize: the Silver award, \$3\,000,
went to 
\begin{itemize}
\item 
Stephen Kelly      (PhD student at Dalhousie University, Canada) 
and
Malcolm I. Heywood (Dalhousie University) 
\end{itemize}
for
``Emergent Solutions to High-Dimensional Multi-Task Reinforcement Learning''
Evolutionary Computation (2018)
\cite{Kelly:2018:EC}.

\item Finally the judges decided to split the 
third prize, the Bronze award,
evenly between two finalists
(\$1,000 each):

\begin{itemize}
\item 
\begin{itemize}
\item 
Emma Hart        (Napier University, Edinburgh, UK)
\item 
Kevin Sim        (Napier University),
\item 
Barry Gardiner   (INRA, Bordeaux, France)
and
\item 
Kana Kamimura    (Shinshu University, Japan)
\end{itemize}
for 
``A Hybrid Method for Feature Construction and Selection to Improve
Wind-damage Prediction in the Forestry Sector''
(GECCO 2017)
\cite{Hart:2017:GECCO}

\item
\begin{itemize}
\item 
Milan Ceska      (Brno University of Technology, Czech Republic),
\item 
Jiri Matyas      (PhD student at Brno University of Technology),
\item 
Vojtech Mrazek   (PhD student at Brno University of Technology),
\item 
Lukas Sekanina   (Brno University of Technology),
\item 
Zdenek Vasicek   (Brno University of Technology),
and 
\item 
Tomas Vojnar     (Brno University of Technology)
\end{itemize}
for 
``Approximating Complex Arithmetic Circuits with Formal Error
Guarantees: 32-bit Multipliers Accomplished''
Proceedings of International Conference On Computer Aided Design
(ICCAD 2017)
\cite{Ceska:2017:ICCAD}.
\end{itemize}
\end{itemize}

\noindent
Details of all the Humie entries can be found on
\href{http://www.human-competitive.org/awards}
     {http://www.human-competitive.\allowbreak{}org/\allowbreak{}awards}

\subsection{Winner}

\begin{figure}%[!p]
\centering
\includegraphics[width=\textwidth]{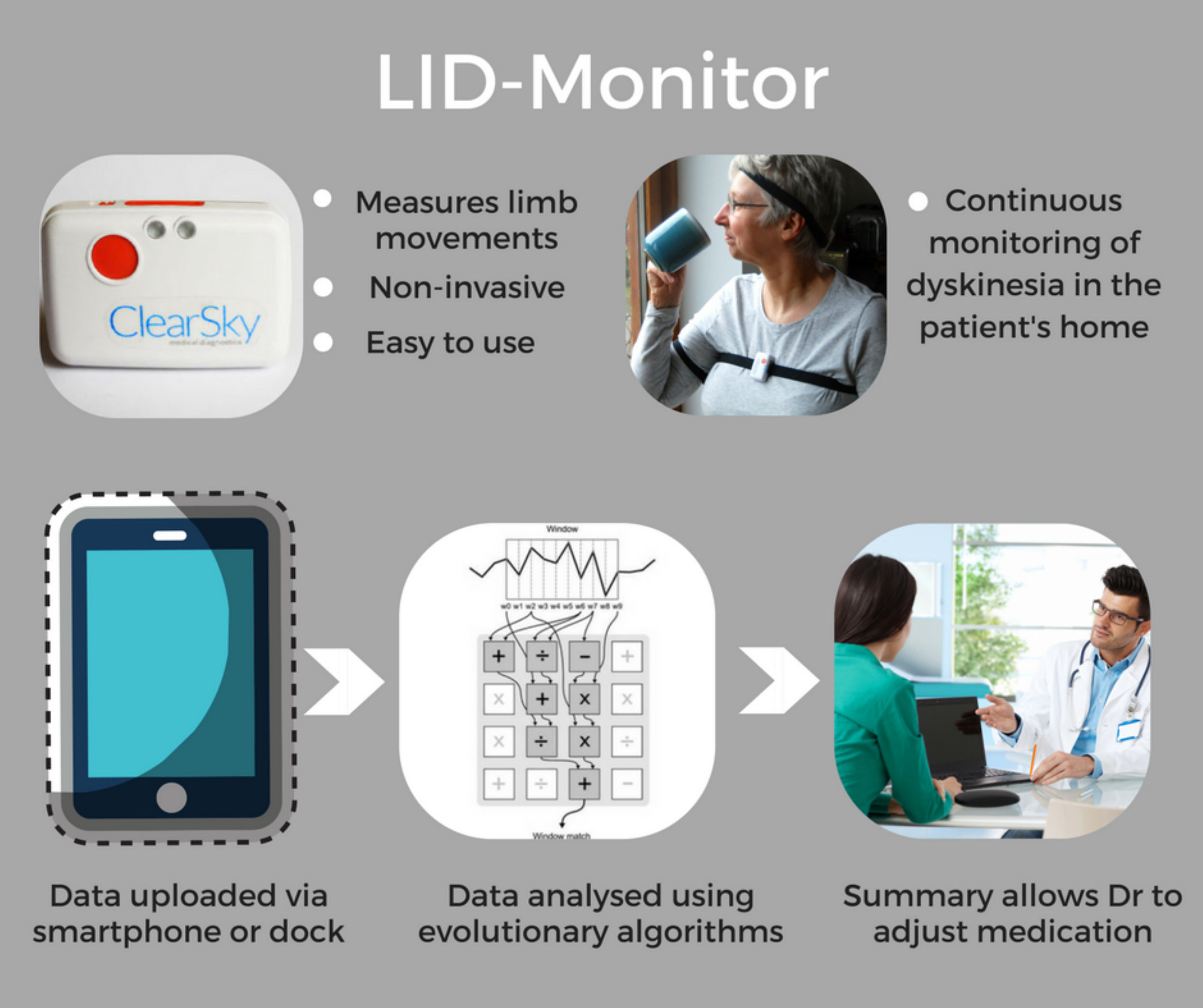}
\caption{Schematic of ClearSky's Parkinson's Disease system which
won the Gold Humie at GECCO-2018.
Monitoring Levodopa--induced dyskinesia (LID) 
gives the patient's doctor
information about the severity and frequency of
patient shaking (dyskinesia) events
and so allows them to better adjust the dosage of Levodopa.
Top left: the mobile monitor and logging system.
These are worn at home by Parkinson's patient (Top Right).
Bottom Centre: 
the logged data are analysed by an algorithm evolved using
Cartesian Genetic Programming and the results
are presented to the patient's doctor (Bottom Right).
}
\label{fig:solution-lid-monitor2}
\end{figure}

Parkinson's Dyskinesia
is an incurable
severe form of Parkinson's disease where the patient
suffers from involuntary jerking movements and muscle spasms (dyskinesia).
However it can be treated
with drugs, e.g.\ Levodopa.
Lones et al.~\cite{Lones:2017:JMS}
were awarded first prize for the invention of a home based
monitoring device that allows dyskinesia to be measured as a patient
goes about their daily routine
(Figure~\ref{fig:solution-lid-monitor2}).
The degree of shaking helps doctors to recommend drug dosage.
The patented monitor uses a predictive model which was trained using
Cartesian Genetic Programming~\cite{Miller:2015:GECCOcomp}.
It has been approved for European clinical use
and is already in routine use internationally
(three large UK Hospitals, 
Leeds, Harrogate and Scarborough)
and one in China (Ruijin Hospital, Shanghai).

To paraphrase the winning entry, they showed the
successful application of
evolutionary computing (CGP)
to resolve a challenging and
life-affecting clinical condition (i.e.\ Parkinson's dyskinesia).
They have published a health economic assessment 
\cite{FILBY2015A358}
which shown that not only will the
introduction of the technology significantly improve the quality of
life but also has the potential
to save the UK's National Health Service over \pounds 84m per year.

\subsection{Runner up}

Google's successfully application of Deep Learning 
on highly parallel hardware (GPUs)
to games,
particularly GO,
is well known.
More recently they have have success with applying it to play computer
based video games
using only the visual cues available to human players
(i.e.~the pixels on the screen).
A particular benchmark is the Atari 2600 suite of games~\cite{nature14236}.

To paraphrase
Kelly and Heywood,
they
applied Genetic Programming~\cite{koza:book}
and were able to 
match the quality of deep learning
but their evolved model was at least three orders of magnitude smaller
allowing real time performance without 
specialized hardware support.
This means the evolved solutions execute on a laptop computer faster than
any form of solution employing Deep Learning.

\subsection{Bronze Prize: Storm Damage to Forests}

Storms can cause severe damage to trees.
For example, in 2009,
a storm lead to losses of $\approx\!\!1.8\,10^{9}$ euros 
to forests in south-west France.
Genetic Programming~\cite{koza:book} gave much better models 
of storm damage, giving better predictions of the causes leading to damage and
so leading to improved forest management
\cite{hart:2018:stormAI}.

\subsection{Bronze Prize: Evolving an Approximate 32-bit multiplier}

Many very clever people have worked on digital electronic circuits 
to do arithmetic for many years.
In just a few hours,
using Cartesian Genetic Programming~\cite{Miller:2015:GECCOcomp},
Milan Ceska et al.\
were able to evolve arithmetic circuits
(e.g.\ for addition or multiplication)
which trade-off circuit size versus accuracy
in a principled way,
with formal guarantees on the maximum permitted error.

\section{Next year}
The $16^{\rm th}$ Humie awards will be held next year
together with
\href{http://gecco-2019.sigevo.org/}
{GECCO}
in Prague, the capital of the Czech Republic,
13-17 July 2019.
Do not forget to enter the competition
\href{http://www.human-competitive.org}
     {http://www.human-competitive.org}

\begin{figure}%[!p]
\centering
\href{http://www.human-competitive.org/}
{\includegraphics[width=\textwidth]{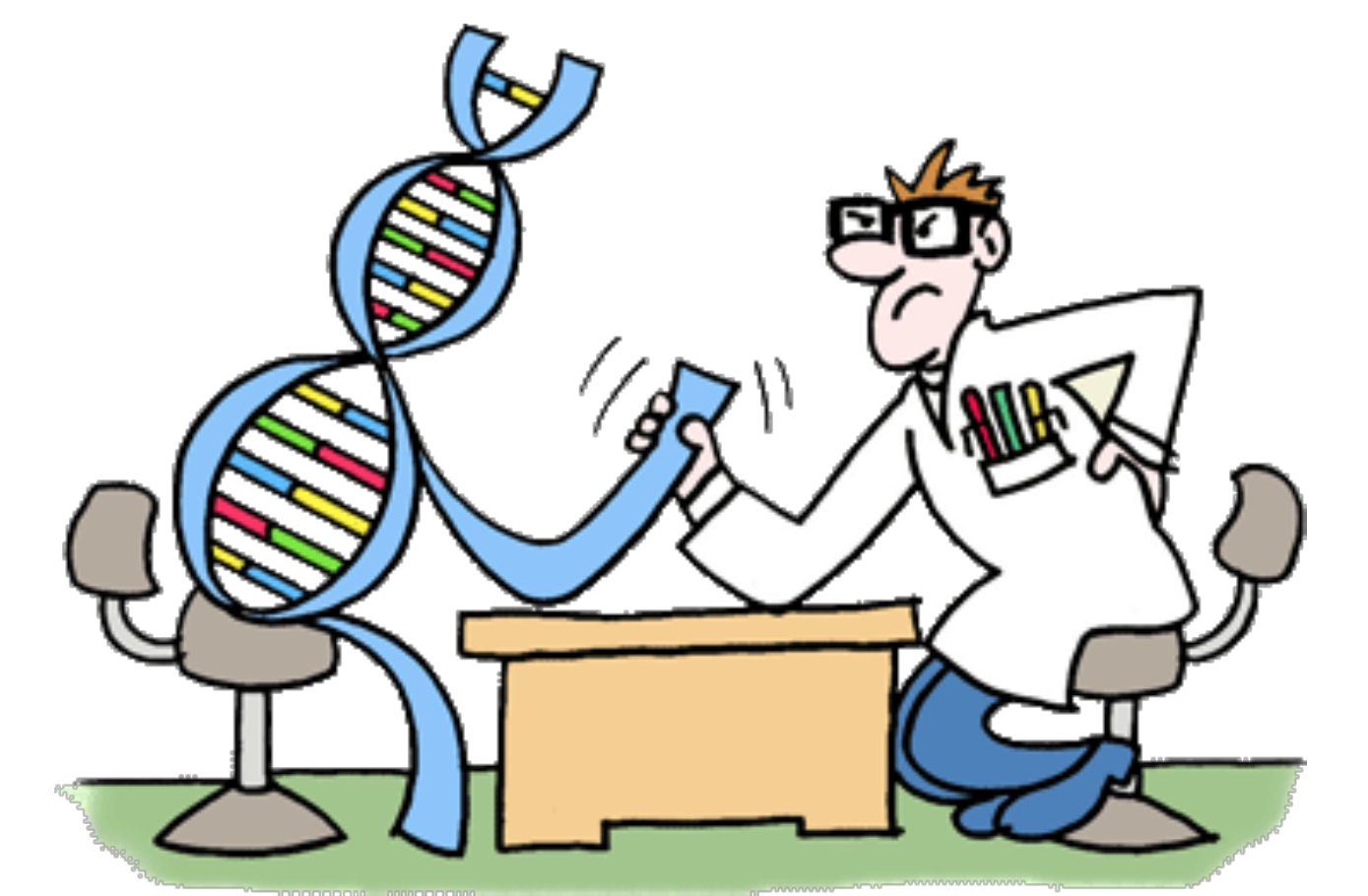}}
\caption{
The Humies logo shows evolutionary computation,
in the shape of an artificial intelligence double spiral DNA,
arm wrestling a human expert
(shown as a white coated ``boffin'').
The goal of the ``Humies'' Awards is to show case 
human competitive results, 
produced by genetic or evolutionary computation
}
\label{fig:humie_logo}
\end{figure}

\clearpage
\newpage
\bibliographystyle{unsrt} %abbrv %named acm
\bibliography{references,gp-bibliography}

\end{document}